\documentclass[onecolumn, amssymb, showpacs]{revtex4-1}				  

\usepackage{graphicx}			
\usepackage{dcolumn}			
\usepackage{bm}					
\usepackage{float}			
\usepackage{amsmath,amsfonts}	
\usepackage{url}					
\usepackage{setspace}		
\usepackage{lineno}   			

\usepackage[colorinlistoftodos]{todonotes}

\begin{document}
 
\begin{space}      	


\title[Fluid-loaded metasurfaces]{Fluid-loaded metasurfaces}      

\author{E.~A. Skelton}			
\author{R.~V. Craster}			
\author{A. Colombi}
\affiliation{Department of Mathematics,  Imperial College London,
  London, SW7 2AZ, U.K.}
\author{D.~J. Colquitt}		
\affiliation{Department of Mathematical Sciences, University of
  Liverpool, Liverpool, L69 7ZL, U.K.}

\date{\today} 
 
\begin{abstract}
  We consider wave propagation along fluid-loaded structures which
take the form of an elastic plate augmented by an array of resonators
forming a metasurface, that is, a surface structured with
sub-wavelength resonators. Such surfaces have had considerable recent
success for the control of wave propagation in electromagnetism and
acoustics, by combining the vision of sub-wavelength wave
manipulation, with the design, fabrication and size advantages
associated with surface excitation. We explore one aspect of recent
interest in this field: graded metasurfaces, but within the context of
fluid-loaded structures.
 
 Graded metasurfaces allow for selective spatial frequency separation
and are often referred to as exhibiting rainbow trapping. Experiments,
and theory, have been developed for acoustic, electromagnetic, and
even elastic, rainbow devices but this has not been approached for
fluid-loaded structures that support surface waves coupled with the
acoustic field in a bulk fluid. This surface wave, coupled with the
fluid, can be used to create an additional effect by designing a
metasurface to mode convert from surface to bulk waves. We demonstrate
that sub-wavelength control is possible and that one can create both
rainbow trapping and mode conversion phenomena for a fluid-loaded
elastic plate model.

\end{abstract}


\maketitle



\section{\label{sec:1} Introduction}

The interaction of sound with compliant structures is a rich area of
research that has generated significant interest, academic and
otherwise, over the years with, at least, five
monographs~\cite{Cremer1988structure,Fahy1987sound,Skelton1997theoretical,Junger1972sound,Crighton1996modern}
on the subject.
The interaction between acoustical and structural waves covers a very
broad range of topics, with an equally broad range of techniques; from
formal mathematical analyses (see, for
example,~\cite{Crighton1996modern}) to more applied research concerned
with submarine acoustics~\cite{Caresta2010}.
In terms of physical problems, there are obvious applications,
such as the generation of noise by flows past elastic structures (see,
for example,~\cite{Howe2008}), flow in pipes~\cite{Yu2013}, scattering
of water waves by flexible ice sheets~\cite{Balmforth1999,Porter2007}, and the
insonification of underwater elastic shells~\cite{Gaunaurd1990}
amongst many others. The field
itself 
has an illustrious history: The  
 overview of the influence of fluid-loading on vibrating
structures provided by Crighton's 1998 Rayleigh Medal
lecture~\cite{Crighton1989} notes that it is almost impossible to find
an area of research in wave motion where Lord Rayleigh has not worked
and that fluid-loaded structures are no exception: 
In particular, Lord Rayleigh analysed the behaviour of air when
excited by a vibrating circular plate~\cite{strutt1896theory} and 
 his contemporaries, such as Lamb, also investigated similar
problems~\cite{Lamb1920}. 

Our interest here stems from recent advances
in so-called metamaterials \cite{pendry96a,pendry99a} that are
man-made composite structures that owe their properties to
sub-wavelength structuration. These metamaterials can be 
designed to have effective properties not found in nature such as
negative refractive index \cite{veselago,pendry2} and these ideas which have proved highly
effective in other areas of wave physics, primarily in electromagnetism \cite{pendry23062006}, in terms of controlling wave
motion but which are now influencing emergent areas such as acoustic
\cite{Liu,craster2012} and seismic \cite{brule,colombi16a,daniel17a}
metamaterials. It is natural therefore to turn to fluid-loaded
structures and evaluate whether metamaterials can be employed to any
effect in this area. Since the structural acoustics of fluid-loading
is inherently connected with a compliant surface such as an elastic
plate or shell it is natural to consider metasurfaces rather than bulk
metamaterials in this context. 

An active area of research in metasurfaces \cite{stru_surf} focusses upon graded
resonator metasurfaces, where
the general concept is that for a graded surface, or waveguide, different wavelengths are trapped at different
spatial positions.
Sub-wavelength microstructures are commonly employed in two ways.
The first approach is to create effective macroscopic wavespeeds that vary spatially, thus achieving the required control of wave propagation.
The second approach involves using deep sub-wavelength resonances to obtained the desired effects.
Whilst the latter approach is significantly more difficult to create, it is far more powerful; hence our aim in the present paper is to extend this concept to fluid-loaded compliant structures.
These ideas are being widely adopted in photonics and phononics due to their
excellent abilities to control, manipulate and filter waves in compact
devices.
Graded and chirped designs include: trapping in rainbow devices
\cite{hess2007,jang11a}, flat focussing mirrors
and lenses in optics, plasmonics and acoustics
\cite{yu14a,maigyte13a,cheng14a,kadic11a,christensen12a,lens_nature}, gradient index lens for acoustic
and flexural waves focussing \cite{martin10a,andrea6}, acoustic
absorbers \cite{acou_rain,jimenez16a} and sound enhancement
\cite{romero13a}.

In the absence of the fluid-loading, it is only very recently that
graded sub-wavelength structures 
 for thin elastic plates has been considered as a chirped graded array
\cite{tian17a}, thin beams \cite{celli1} or in the context of gradient
index (GRIN) lenses created by graded structuration to control elastic
symmetric ($S_0$) and antisymmetric ($A_0$) waves
\cite{torrent_plate2} and to obtain deeply sub-wavelength focussing
\cite{andrea4},  cloaking \cite{andrea5} and GRIN lenses
(e.g. Luneburg, Maxwell-fisheye and Eaton lenses 
\cite{luneburg_optic,celli1}) using resonator arrays \cite{andrea6}. 

These are all primarily for effectively scalar wave systems and do not
have the degrees of freedom necessary to demonstrate mode conversion from surface
to bulk waves and so can only demonstrate trapped rainbow phenomena.
  More recently arrays of resonators atop elastic
half-spaces have demonstrated, theoretically, numerically \cite{colombi16a,daniel17a} and
experimentally in ultrasonics \cite{colombi17a}, that one can 
 generate rainbow trapping phenomena and additionally manipulate surface
Rayleigh waves and mode convert between Rayleigh and bulk shear
waves. This mode conversion creates a surface device that is
simultaneously reflectionless, yet has zero transmission into surface
waves, and moreover that this occurs over an ultra-broadband range of
frequencies. The rainbow trapping also has important implications as
one can strongly enhance the surface wave amplitude at particular
points along the interface and this is selectively achieved by
altering the frequency.

\begin{figure}  
\vspace{0cm}\hspace{0cm}\includegraphics[width=9cm]{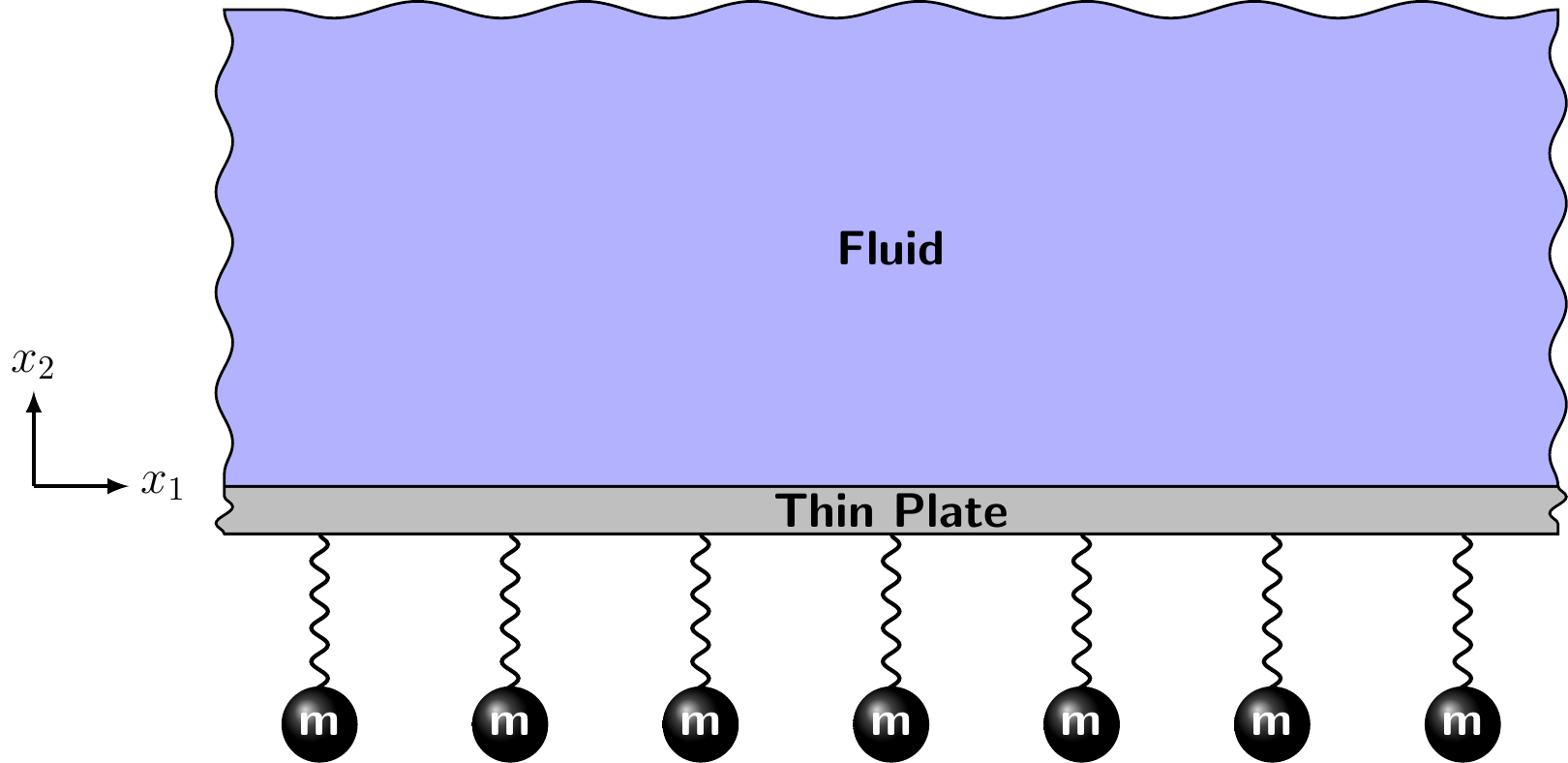}
\caption{A fluid-loaded elastic plate augmented by a periodic array of
identical mass-spring resonators.}
\label{fig:plate}
\end{figure}

The ability in elasticity to selectively use sub-wavelength graded resonator systems to
influence both surface and bulk waves, as well as the coupling between
them, 
motivates us to consider fluid-loaded structures. Although the governing equations are
scalar, there is a coupling between the elastic waves in the plate and the acoustic pressure field in the fluid;
it is this coupling that allows the structure to support a subsonic surface wave in addition to the bulk acoustic wave. We begin, in
section \ref{sec:plates}, by
assessing the influence of attaching periodically arranged identical resonators, modelled
 by the simplest system, that of a mass-spring, to the
 fluid-loaded structure (as shown in Fig. \ref{fig:plate}). Scattering and radiation by periodic systems in the
 fluid-loaded literature are natural models of ribbed and reinforced
 structures such as a ship's hull and have been much studied
 \cite{evseev1973,Leppington1978,Mace1980a,Skelton1990a,pierce,weaver_sprung} amongst many others, and
 these analyses can be adapted to incorporate the resonator array. We deduce
 the dispersion relations that characterise the behaviour of waves
 within this periodic fluid-loaded plate- resonator system. These are 
 utilised in section \ref{sec:results} to interpret the
 influence of grading the resonators by, for instance, increasing or
 decreasing the masses spatially along the array (as shown in
 Fig. \ref{fig:plate-graded}) and this interpretation is complemented
 by careful numerical simulations for an incoming subsonic surface
 plate wave incident upon graded arrays. Given the interest in underwater
 acoustics and maritime structures we choose our numerical examples to
 have aluminium as the plate (thickness $h=0.01$~m, Young's modulus 
 $E_p=69\times 10^9$~N/m$^2$, Poisson
 ratio $\nu_p=0.334$ and density $\rho_p=2700$~kg/m$^3$) and water
 (density $\rho_f=1000$~kg/m$^3$, and sound speed $c_f=1500$~m/s) as the fluid. 
  Finally, we draw together some
 comments and future perspectives on the use of metamaterial ideas for
 fluid-loaded structures. 

\begin{figure}  
\vspace{0cm}\hspace{0cm}\includegraphics[width=9cm]{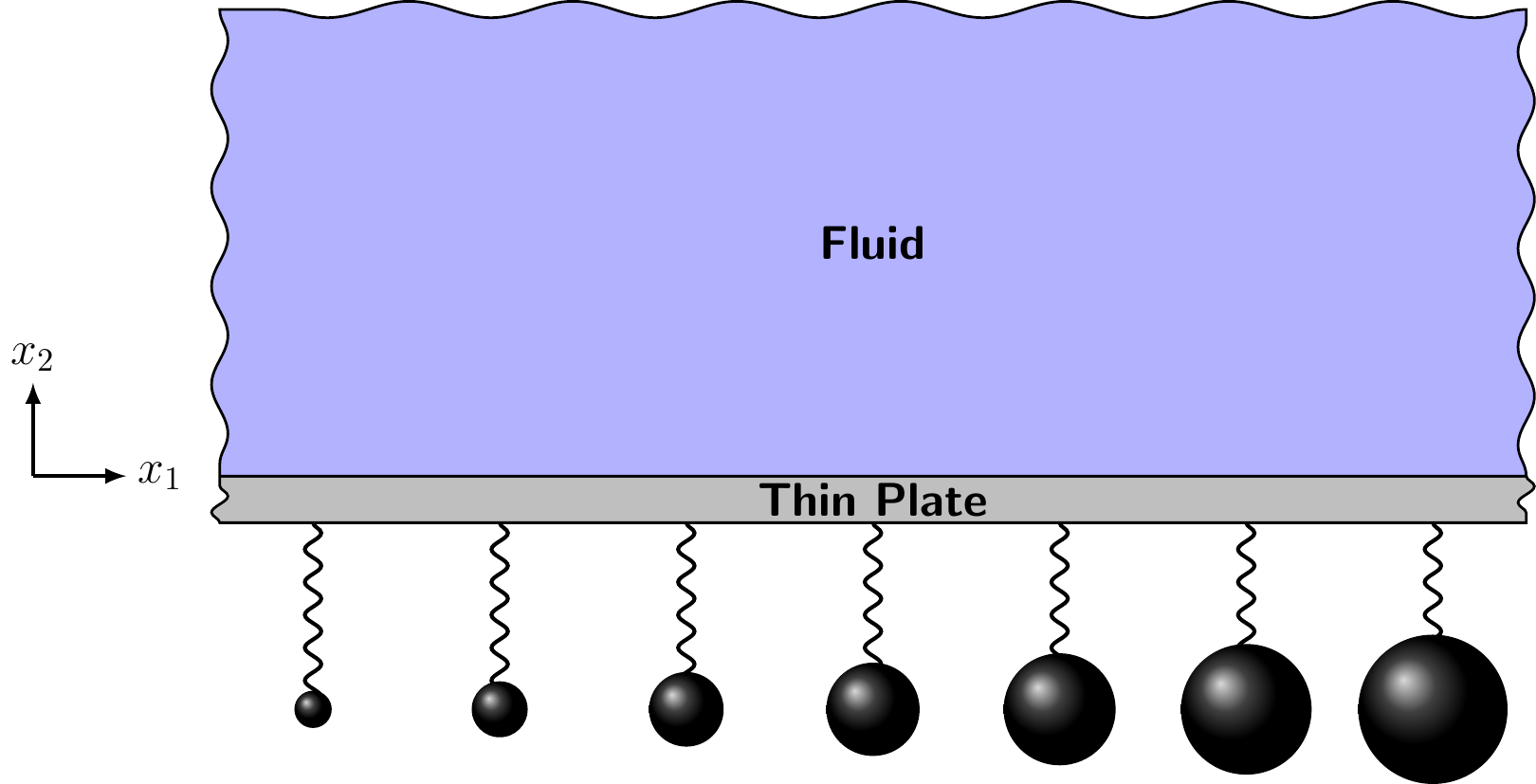}
\caption{A fluid-loaded elastic plate augmented by a periodic array of
graded mass-spring resonators.}
\label{fig:plate-graded}
\end{figure}

\section{Fluid-loaded thin elastic plate - resonator
  arrays \label{sec:plates}}

\subsection{Formulation}

We consider a thin elastic plate governed by the Kirchhoff-Love plate
equations, as the simplest model of an elastic plate that captures the
relevant physics for underwater acoustics \cite{Junger1972sound} whilst
recognising that this model can be complexified to allow the plate to
be either anisotropic or thicker such that, say, the Mindlin equations become
relevant; these generalisations are readily incorporated. We also
restrict our attention to two-dimensions for clarity and as it will
allow us to illustrate the key concepts without being over-burdened by
excessive algebra. 

Here, a thin elastic plate of thickness $h$, density $\rho_p$, Young's
modulus $E_{p}$ and Poisson's ratio $\nu_p$ is located at $z=0$,
$-\infty <x<\infty$. It is bounded on its upper surface by an infinite
acoustic fluid ($-\infty <x<\infty$, $0<z<\infty$) of density $\rho_f$
and sound speed $c_f$. An infinite periodic array of mass-spring systems
are attached to the lower surface of the plate, exerting point reaction
forces, as shown in Fig. \ref{fig:plate}. 

The (2D) equation of motion for a thin fluid-loaded plate is written as
\begin{equation}
D\frac{\partial^4\hspace{1ex}}{\partial x^4} u(x,t)
+ M\frac{\partial^2\hspace{1ex}}{\partial t^2}u(x,t) = 
\sum_{n=-\infty}^{\infty} F_{n}\delta (x-nd) - p(x,0,t)
{\rm ,}
\label{eq:plates.1}
\end{equation}
in which $D=E_{p}h^3/12(1-\nu_{p}^{2})$ is the plate bending
stiffness, and $M=\rho_p h$, $u(x,t)$ is the plate displacement,
$p(x,z,t)$ is the acoustic pressure in the fluid 
 \cite{Junger1972sound} and $F_{n}$ is the reaction force of the $n$'th mass-spring system located at $x_{n}=nd$.
The equation satisfied by the pressure in the acoustic fluid, with
sound speed $c_f$, is the wave equation
\begin{equation}
\left( \frac{\partial^2\hspace{1ex}}{\partial x^2}
+\frac{\partial^2\hspace{1ex}}{\partial z^2}
-\frac{1}{c_f^{2}}\frac{\partial^2\hspace{1ex}}{\partial t^2} \right) p(x,z,t) = 0
{\rm ,}\label{eq:plates.2}
\end{equation}
with the continuity condition at the plate surface such that the normal velocity in the fluid there is equal to the normal velocity of the plate.
The reaction forces $F_{n}$ are determined from the equations of motion of the mass-spring system:
\begin{eqnarray}
F_{n} & = & -m\frac{{\rm d}^{2}\hspace{1ex}}{{\rm d}t^2} u_{mn}(t)
{\rm ,}\label{eq:plates.3}\\
F_{n} & = & \lambda (u_{mn}(t)-u(nd,t))
{\rm ,}\label{eq:plates.4}
\end{eqnarray}
where $m$ is the mass, $\lambda$ is the spring constant and $u_{mn}$ is the displacement of the attached mass at position $x_{n}=nd$.

\subsection{Dispersion relation\label{sec:disp}}

It is convenient to first consider time-harmonic motion, in which all
the dependent variables are proportional to $\exp (-{\rm i}\omega
t)$ and henceforth this exponential factor is
suppressed. Thus, equations~(\ref{eq:plates.1})--(\ref{eq:plates.4})
become
\begin{equation}
D\frac{\partial^4\hspace{1ex}}{\partial x^4} u(x)
- M\omega^{2}u(x)  = 
\sum_{n=-\infty}^{\infty} F_{n}\delta (x-nd) - p(x,0)
{\rm ,}\label{eq:disp.1}
\end{equation}

\begin{eqnarray}
0 & = & \left( \frac{\partial^2\hspace{1ex}}{\partial x^2}
+\frac{\partial^2\hspace{1ex}}{\partial z^2}
+k^2  \right) p(x,z)
{\rm ,}\label{eq:disp.2} \\
F_{n} & = & m\omega^{2} u_{mn}
{\rm ,}\label{eq:disp.3}\\
F_{n} & = & \lambda (u_{mn}-u(nd))
{\rm .}\label{eq:disp.4}
\end{eqnarray}
 with $k=\omega/c_f$ being the fluid
wavenumber. 
Eliminating the mass displacements from (\ref{eq:disp.3}) and
(\ref{eq:disp.4}) gives
\begin{equation}
F_{n} = -m\omega^{2}\lambda u(nd) / (m\omega^{2}-\lambda) = -L(\omega )u(nd)
{\rm .}\label{eq:disp.5}
\end{equation}
For the mass-spring system here $L(\omega )=m\omega^{2}\lambda /
(m\omega^{2}-\lambda)$ and resonance occurs at $\omega=\sqrt{\lambda/m}$; more complicated arrangements of
masses and springs, or rods, can be considered leading to more complex functions
encapsulated within $L(\omega)$.
Fourier transform pairs for the plate displacement and the acoustic
 pressure are introduced, for example, for the plate displacement:
\begin{eqnarray}
u(x) &= & \frac{1}{2\pi}\int_{-\infty}^{\infty} \bar{u}(\alpha)\exp ({\rm i}\alpha x){\rm d}\alpha{\rm,}
\label{eq:disp.6}\\
\bar{u}(\alpha) &= & \int_{-\infty}^{\infty} u(x)\exp (-{\rm i}\alpha x){\rm d}x{\rm.}
\label{eq:disp.7}
\end{eqnarray}
Transforming (\ref{eq:disp.6}) we see the pressure is
\begin{equation}
\bar{p}(\alpha,z) = \bar{p}(\alpha,0){\rm e}^{{\rm i}\gamma z}
\label{eq:disp.8}
\end{equation}
where $\gamma=\sqrt{k^{2}-\alpha^{2}}$ and the continuity condition at
the plate surface couples the pressure to the plate displacement and 
\begin{equation}
{\rm i}\gamma \bar{p}(\alpha, 0)=\rho\omega^{2}\bar{u}(\alpha)
{\rm .}\label{eq:disp.9}
\end{equation}

Making use of the Poisson Summation Formula~\cite{deitmar2014principles}, the reaction force terms of equation~(\ref{eq:disp.1}) can be written in the form
\begin{equation}
\sum_{n=-\infty}^{\infty}F_{n}\delta (x-nd) =
\frac{-L(\omega)}{d}u(x)\sum_{n=-\infty}^{\infty} \exp (2\pi {\rm i}nx/d)
{\rm ,}\label{eq:disp.10}
\end{equation}

 from which the transformed Eq. (\ref{eq:disp.1}) becomes 
\begin{equation}
S(\alpha,\omega)\bar{u}(\alpha)+\frac{L(\omega)}{d}\sum_{n=-\infty}^{\infty}\bar{u}(\alpha -2n\pi/d) =0
{\rm ,}\label{eq:disp.13}
\end{equation}
where $S(\alpha,\omega)=D\alpha^4-\omega^2M-{\rm
  i}\rho\omega^{2}/\gamma$. 

In the absence of any forcing or resonators, that is if we consider a pristine
elastic plate with fluid loading, the dispersion relation is given by
$S(\alpha,\omega)=0$. Going one level even simpler, that is to an
elastic plate {\it in vacuo} $D\alpha^4-\omega^2M=0$ and
$\alpha=(M\omega^2/D)^{1/4}$ giving the flexural plate wavenumber-frequency relationship that is nonlinear and which indicates that
these plate waves are dispersive. The addition of the fluid coupling
leads to these flexural plate waves acquiring a complex component and
they become leaky waves, and we shall not consider them further here. 
The fluid-loaded system also allows for a
subsonic (relative to the overlying fluid) surface wave that exists
due to the coupling between the fluid and plate and which
exponentially decays into the fluid. This wave is effectively the
primary excitation of the arrays we describe and our aim is to
investigate how this surface wave interacts with the resonant array.

Returning to the array of resonators, Eq. \eqref{eq:disp.13} becomes
\begin{equation}
\left[ 1+
\frac{L(\omega)}{d}\sum_{n=-\infty}^{\infty}\frac{1}{S(\alpha -2n\pi/d,\omega)}
\right]
\sum_{n=-\infty}^{\infty}\bar{u}(\alpha -2n\pi/d)=0
\label{eq:disp.14}
\end{equation}
and the dispersion relation for the periodic system is therefore
\begin{equation}
1+\frac{L(\omega)}{d}\sum_{n=-\infty}^{\infty}\frac{1}{S(\alpha -2n\pi/d,\omega)}=0
{\rm .}\label{eq:disp.15}
\end{equation}
There are limits of (\ref{eq:disp.15}) that are of interest and which
can be approached analytically. For no fluid-loading the infinite sum of (\ref{eq:disp.15}) can be
evaluated exactly, as in for example~\cite{daniel17a}, allowing a
closed form expression for the dispersion relation. With fluid present
in the upper half-space the situation is more complicated and a closed
form expression appears unavailable. Leading order asymptotic approximations for sums of this form are available~\cite{Skelton1990a} in the heavy fluid-loading limit for which $\tilde{\alpha}_{1}>>k$, and $\tilde{\alpha}_{1}>>\tilde{\alpha}_{0}$,
\begin{equation}
\sum_{n=-\infty}^{\infty}\frac{1}{S(\alpha -2n\pi/d,\omega)} \sim 
\frac{-d}{5D\tilde{\alpha}_{1}^{3}}\left\{ \cot\left(\frac{2\pi}{5}\right)  \right.
\left. +
\frac{\sin(\tilde{\alpha}_{1}d)}{\cos(\alpha d)-\cos(\tilde{\alpha}_{1}d)}\right\}
{\rm ,}\label{eq:DispHFL.1}
\end{equation}
in which $\tilde{\alpha}_{1}=(\rho\omega^{2}/D)^{\frac{1}{5}}$ is the wavenumber in the heavily fluid-loaded plate
and $\tilde{\alpha}_{0}=(\omega^2M/D)^{\frac{1}{4}}$, the wavenumber of the \emph{in vacuo} plate.
Hence, in this heavy fluid-loading approximation,  we rewrite $\alpha$ as a function of $\omega$ as
\begin{equation}
\cos(\alpha d) = \cos (\tilde{\alpha}_{1}d) + \frac{\sin(\tilde{\alpha}_{1}d)}{
\frac{5D\tilde{\alpha}_{1}^{3}}{L(\omega)} -\cot\left(\frac{2\pi}{5}\right)}
{\rm .}\label{eq:DispHFL.2}
\end{equation}
From this expression, for a given value of $\omega$, real values of
$\alpha$ are only obtained 
 when the absolute value of the right hand side of~(\ref{eq:DispHFL.2})
is less than or equal to 1. 
 Numerical examination of this dispersion relation shows that in this
low frequency, heavy fluid-loading, asymptotic regime the dispersion
diagram does not exhibit the band gaps necessary to provide
interesting metamaterial properties.  Thus, we proceed to investigate the full dispersion relation, away from the aforementioned asymptotic limits, numerically. 

\begin{figure}  
\includegraphics[width=10cm]{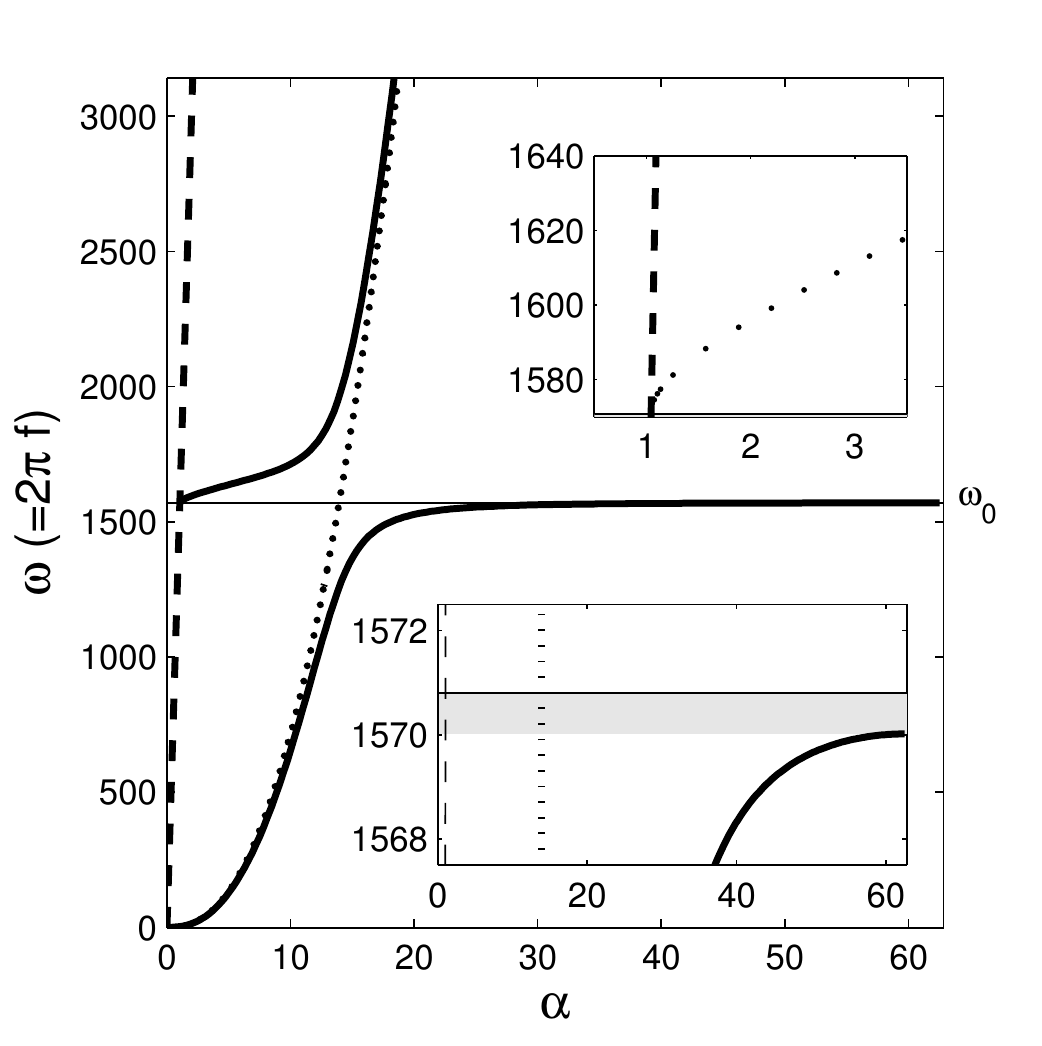}
\caption{Dispersion plot showing frequency $\omega = 2\pi f$ as a
  function of horizontal wavenumber $\alpha$  (solid black lines) for
  1~cm aluminium plate in water and periodic mass-spring resonators
  with spacing $d=5$~cm. Range for $\alpha$ is $[
  0,\pi/d]$. Asymptotes shown: dashed line is $\alpha =\omega/c_f$
  (sound line),  dotted line is unconstrained fluid-loaded plate
  dispersion plot showing $\tilde \alpha_1$, solid black line is resonance frequency of resonators. Insets show zoom-in on very narrow band gap (shaded grey), and zoom-in on the asymptote to the sound line.}
\label{fig:AlphaDisp.1}
\end{figure}


Returning to~(\ref{eq:disp.15}), and writing it as
\begin{equation}
\sum_{n=-\infty}^{\infty}\frac{1}{S(\alpha_n,\omega)}=\frac{-d}{L(\omega)}
\label{eq:DispNum.1}
\end{equation}
in which $\alpha_{n}=\alpha-2n\pi /d$, we note that the right hand side is always real, whereas the left hand side will have an imaginary part unless $\gamma_{n}=\sqrt{k^2-\alpha_{n}^{2}}$ is imaginary for each value of $n$. Thus, for any given value of $\omega$, this restricts the possible values of $\alpha$:
\( k<\alpha<2\pi /d-k\) 
when 
\(
0<\alpha<2\pi /d\) 
and \(0<k<\pi /d\).

%
\begin{figure}  
\includegraphics[width=8cm]{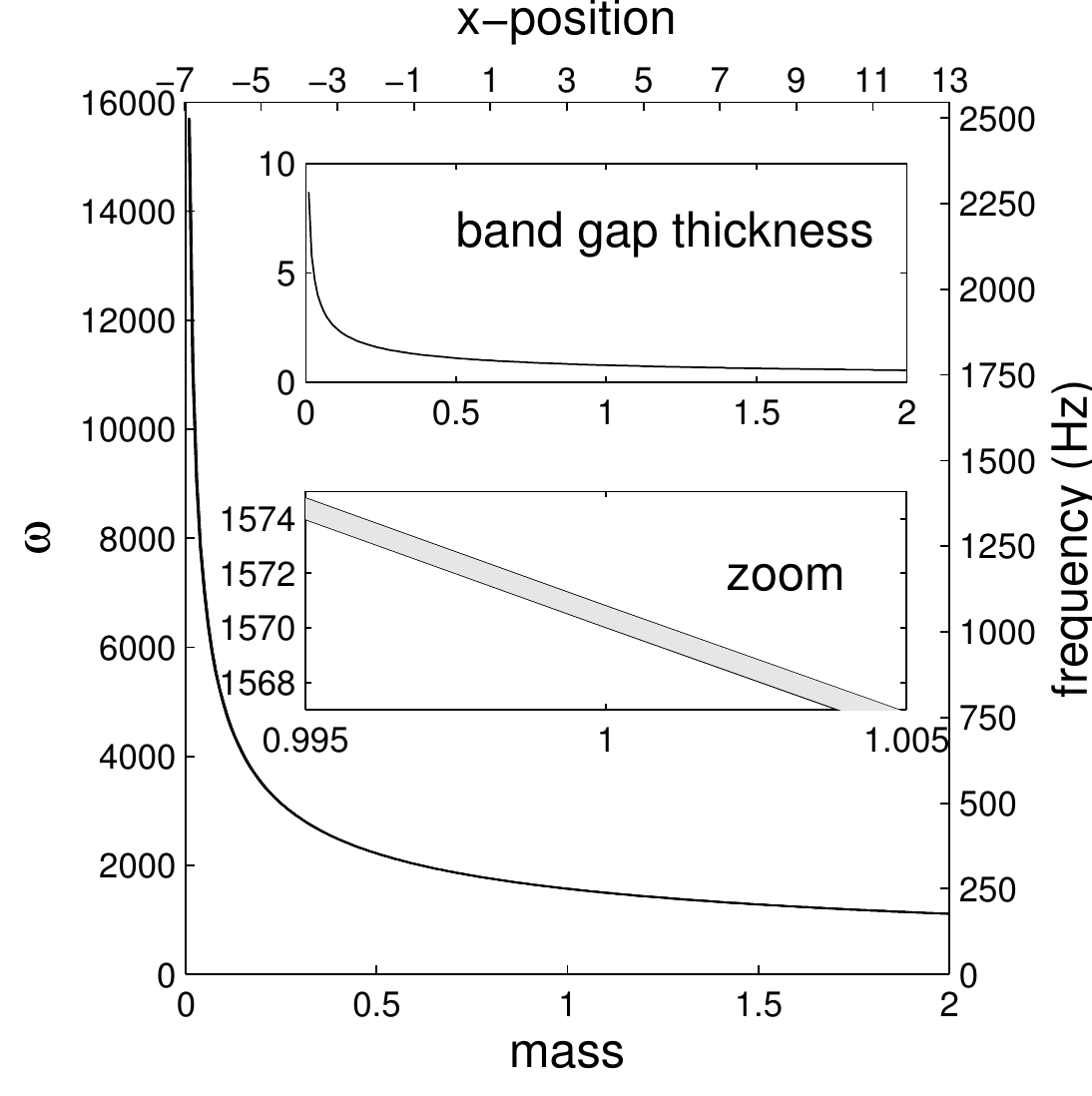}
\caption{Frequency band gap as a function of resonator mass for identical periodically spaced resonators. The upper limit of the band gap is the resonance frequency of the mass-spring system. The band gap too  narrow to be observable on this scale. The lower inset shows a zoom in of the main plot with the shaded area the band gap. The upper inset shows, as a line plot, the band gap width as a function of mass. 
}
\label{fig:BandGap}
\end{figure}

\begin{figure}  
\includegraphics[width=7cm]
{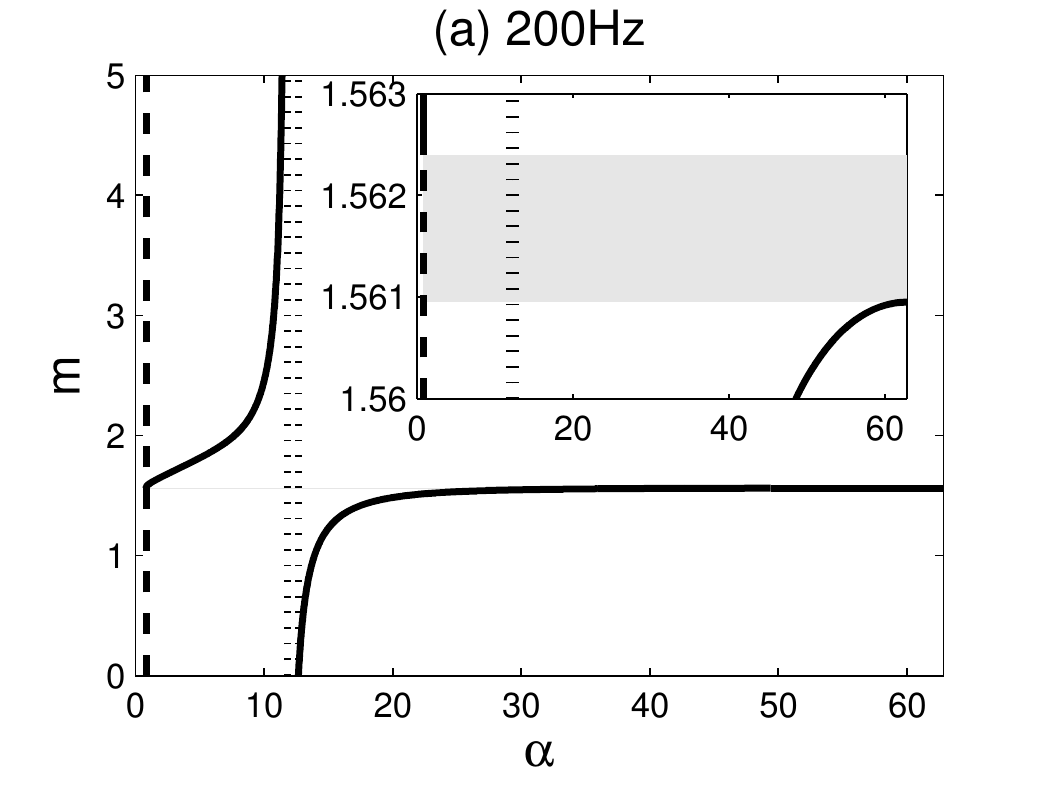}
\includegraphics[width=7cm]
{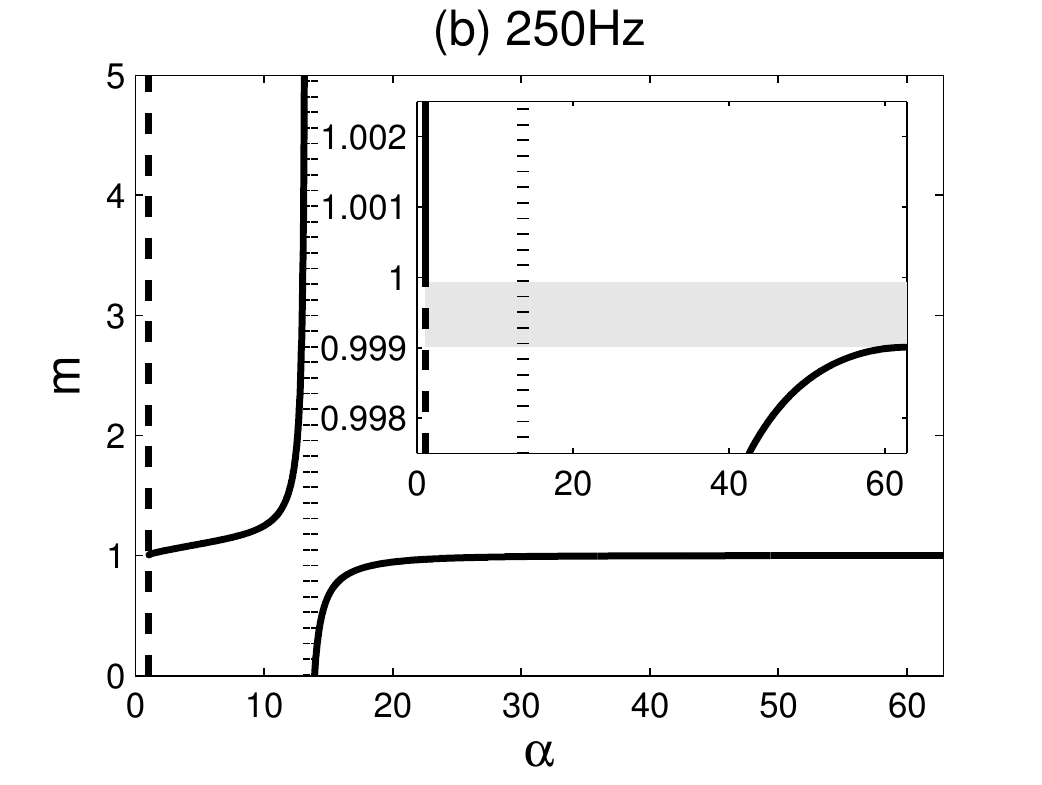}
\includegraphics[width=7cm]
{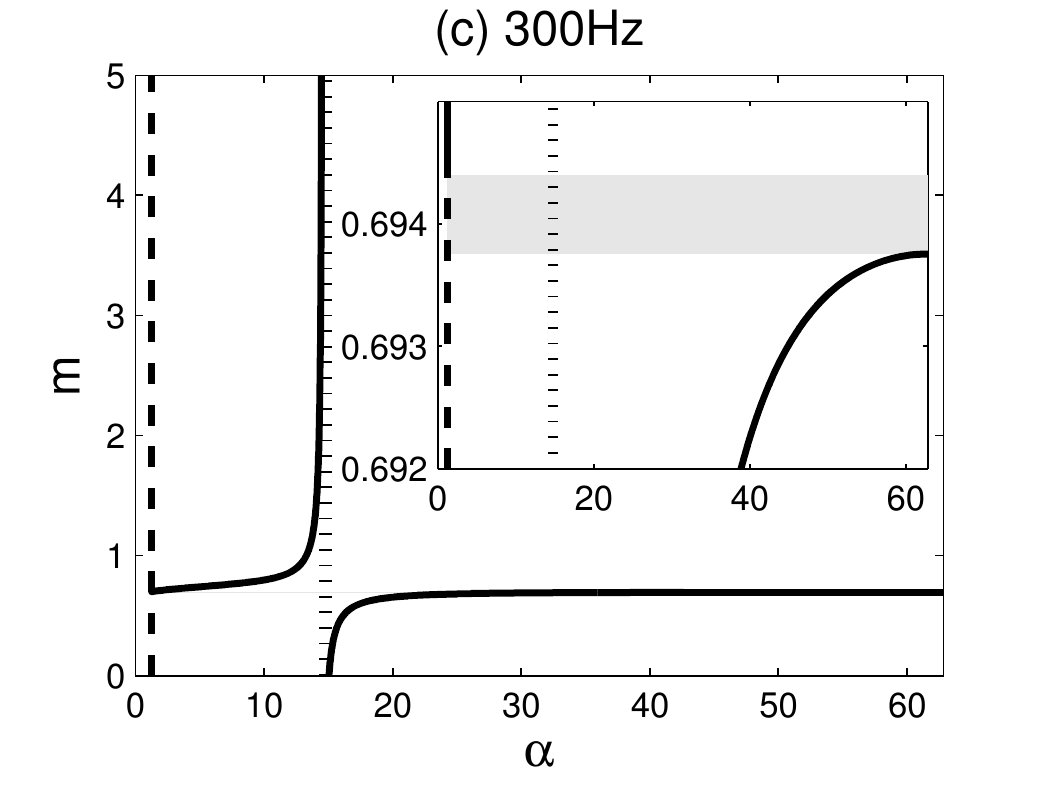}
\caption{For fixed frequencies (200, 250 and 300~Hz from top to
  bottom), spring constant and spacing dispersion plot showing mass as
  a function of horizontal wavenumber for 1~cm aluminium plate in
  water. Asymptotes shown: dashed line is $\alpha =\omega/c_f$ (sound
  line), solid line is obtained numerically, dotted line is the wavenumber of the unconstrained fluid-loaded plate. Inset shows zoom-in on very narrow band gap (shaded grey).}
\label{fig:MassDisp.1}
\vspace{-0.5cm}
\end{figure}

Due to the underlying periodicity, the dispersion diagram repeats in
$\alpha$ with period $2\pi /d$. To find the roots we note that for any
given value of $\omega$, $S(\alpha,\omega)$ has one real positive root and
therefore, $1/S(\alpha,\omega)$ passes through an infinity there and changes
sign. Thus, the left hand side of~(\ref{eq:DispNum.1}), as a function
of $\alpha$, periodically varies from $-\infty$ to $\infty$ and must
therefore equal the right hand side at least once in each such
interval. For heavy fluid loading such solutions are close to the
location of the infinities, but are found numerically by starting the
searches very close to the location of the infinity (where the sign is
known) and looking for a change of sign.

We choose typical parameters in Fig. \ref{fig:AlphaDisp.1} that shows the
fluid-loaded plate - resonator curves together with the resonator
frequency ($\omega_0$), the acoustic sound line $\omega=\alpha c_f$
(dashed line) and the sub-sonic coupled fluid-plate surface wave
(dotted 
line). The effect of the resonator is to punch through the surface
wave mode to create a very narrow band gap, shown as the gray region
in the inset to
Fig. \ref{fig:AlphaDisp.1}, and the hybridisation of the mode. The
hybridisation shows a transition, as one decreases frequency and
approach the band gap from above, in behaviour from the coupled
fluid-solid surface wave to the bulk acoustic mode. This behaviour,
arising from the avoided crossing between the surface wave dispersion
relation and the local resonance is the acoustic manifestation of
a surface plasmon polariton \cite{otto68a} in condensed matter
physics which is one of the main approaches in nanofabricated devices
used to design and control the propagation of light. A key difference
in this fluid-loaded analogy is that the mode below resonance emerges
from a surface wave that is dispersive and so the analogy is not exact.

For typical underwater parameters the band-gap thickness remains
narrow over a wide range of frequencies and resonator mass, see
Fig. \ref{fig:BandGap}, and such a narrow frequency band suggests
that the effect of the band-gap will be only restricted to very
limited frequencies and be of little use. This is relatively common in
graded resonator systems, the resonating rods atop an elastic surface
interacting with surface Rayleigh waves \cite{daniel17a} have
similar behaviour, and to take advantage of the band-gap the key
addition is to make the resonant array spatially dependent in some
way. The spatial dependence can come through a gradual change in the
array spacing, a graduation of spring constant or mass in each
resonator along the array. We choose to alter the masses in the array
by either linearly ramping them up, or down, with distance and illustrate how
this influences an incident surface wave coming from a pristine region
of the plate not containing resonators striking a semi-infinite array
of resonators. The choice of mass change with position mean that the
horizontal axis is equivalently the $x$-position of the resonators and we
note that for a mass of 1~kg the position is $3$~m and the frequency is
250~Hz and these choices are used later. Fig. \ref{fig:BandGap} allows
us to read off the position,
 $x=(\lambda/\omega^2-m(0))/(dm/dx)$, as a function of $m(x)$, 
 at which waves are no longer supported by
the array as a function of position and thus this curve gives the
turning point and thus shows the spatial selection by frequency.

\subsection{Graded array\label{sec:Variable}}

The conventional dispersion diagram of Fig.~\ref{fig:AlphaDisp.1}  for
frequency versus wavenumber comes from Eq.~(\ref{eq:disp.15}), with
$L(\omega)$ defined in~(\ref{eq:disp.5}) and all parameters (plate and
fluid properties, mass-spring properties and the spacing) remaining
constant. An alternative, less common, but instructive diagram are
those 
of Fig. \ref{fig:MassDisp.1} that considers how the modes vary with a parameter,
but now holding frequency fixed. As frequency decreases, from top to
bottom, in  Fig. \ref{fig:MassDisp.1}, one sees the bandgap move
downwards as predicted by Fig. \ref{fig:BandGap} and the vertical
asymptote more to the right, that is, to shorter
waves. Fig. \ref{fig:MassDisp.1} gives a valuable interpretation, for
graded structures, as one can follow a branch from, say, high mass to
low mass or vice-versa, and then immediately see how the wave system
behaves. Before embarking upon this interpretation we extract the
exact mass-frequency relationship. 

The mass and spring constants only enter into~(\ref{eq:disp.15}) via
the $L(\omega)$ term, and the dispersion relation can be re-written as
\begin{equation}
\frac{\omega^{2}-\lambda /m}{\omega^{2}\lambda}=
\frac{-1}{d}\sum_{n=-\infty}^{\infty}\frac{1}{S(\alpha -2n\pi/d,\omega)}
{\rm .}\label{eq:mass.1}
\end{equation}
Hence, if $\omega$ and $\lambda$ are held constant, then the relation
between resonator mass $m$ and wavenumber $\alpha$ is explicit:
\begin{equation}
m =\frac{\lambda}{\omega^{2}\left( 1+\displaystyle{\frac{\lambda}{d}\sum_{n=-\infty}^{\infty}\frac{1}{S(\alpha -2n\pi/d,\omega)}} \right) }
{\rm ,}\label{eq:mass.2}
\end{equation}
 and similarly, holding $\omega$ and $m$ constant, a relation between
 resonator spring constant $\lambda$ and $\alpha$ also emerges:
\begin{equation}
\lambda =\frac{\omega^{2}}{\left( \displaystyle{\frac{1}{m}+\frac{\omega^{2}}{d}\sum_{n=-\infty}^{\infty}\frac{1}{S(\alpha -2n\pi/d,\omega)}} \right) }
{\rm .}\label{eq:mass.3}
\end{equation}

\begin{figure}  
\includegraphics[width=7cm]{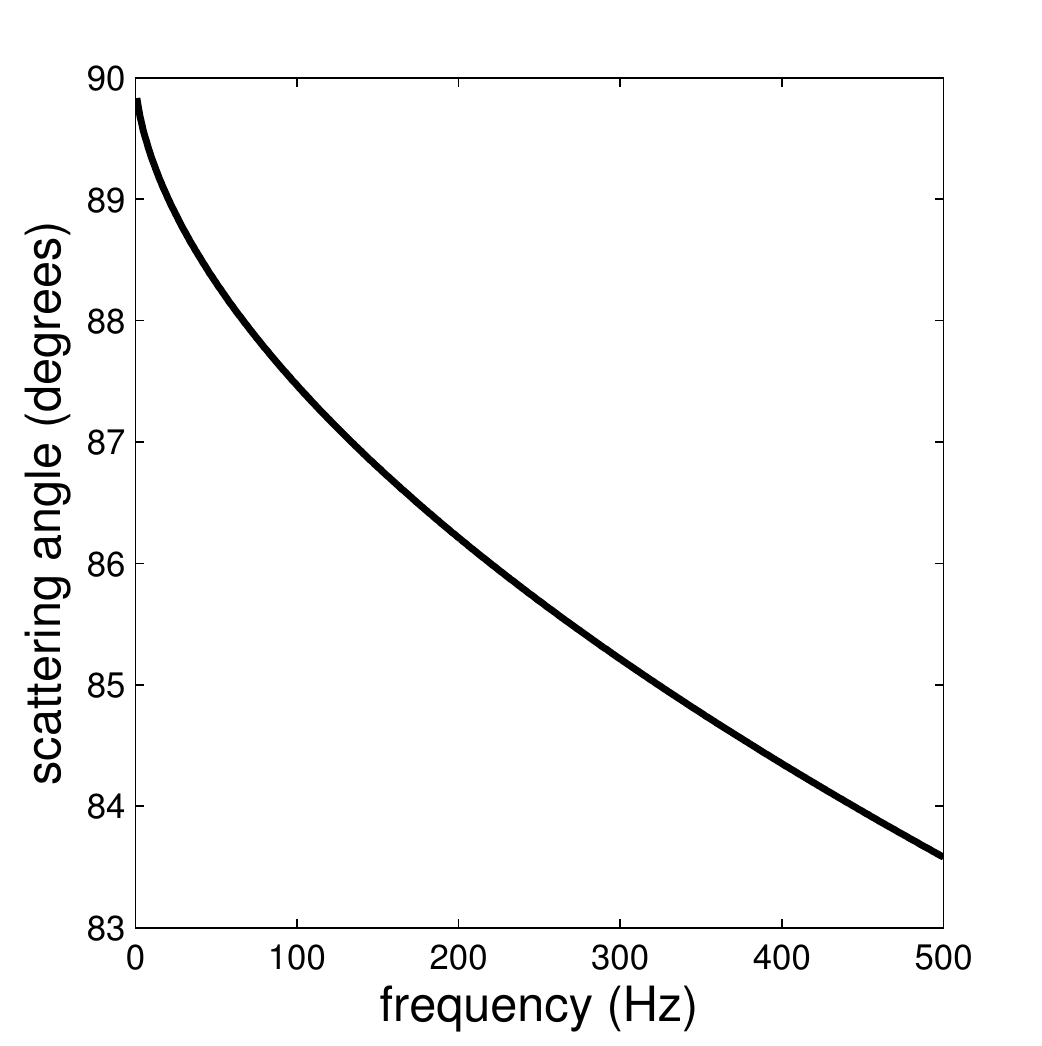}
\caption{The variation of scattering angle with frequency.}
\label{fig:anglefreq}
\end{figure}

Plotting the mass versus wavenumber, using (\ref{eq:mass.2}), in
Fig. \ref{fig:MassDisp.1} allows us to predict and interpret the
behaviour of waves associated with a plate with an array of
mass-spring resonators for which the spacing and spring constant are
held constant, but the mass of the resonators varies (smoothly and
monotonically) with position.

\begin{figure}
\includegraphics[width=10cm]{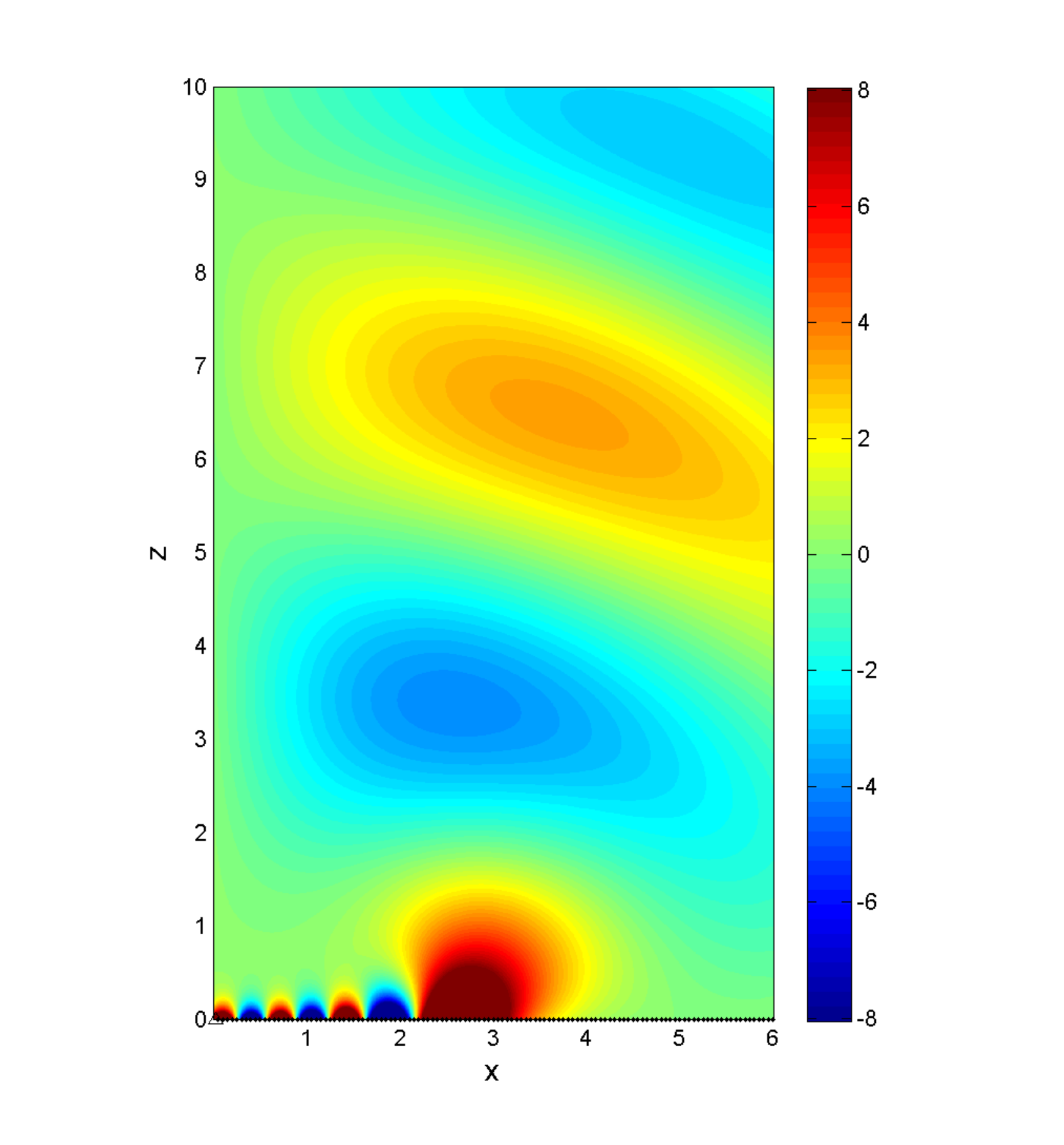}
\caption{Time-harmonic (250Hz) snapshot of pressure in water above 1~cm aluminium plate with mass-spring resonators. Resonator mass decreases with $x$. Resonant frequency of mass-spring located at $x=3$ is 250Hz.}
\label{fig:RampDownPress}
\end{figure}
\begin{figure}  
\includegraphics[width=8cm]
{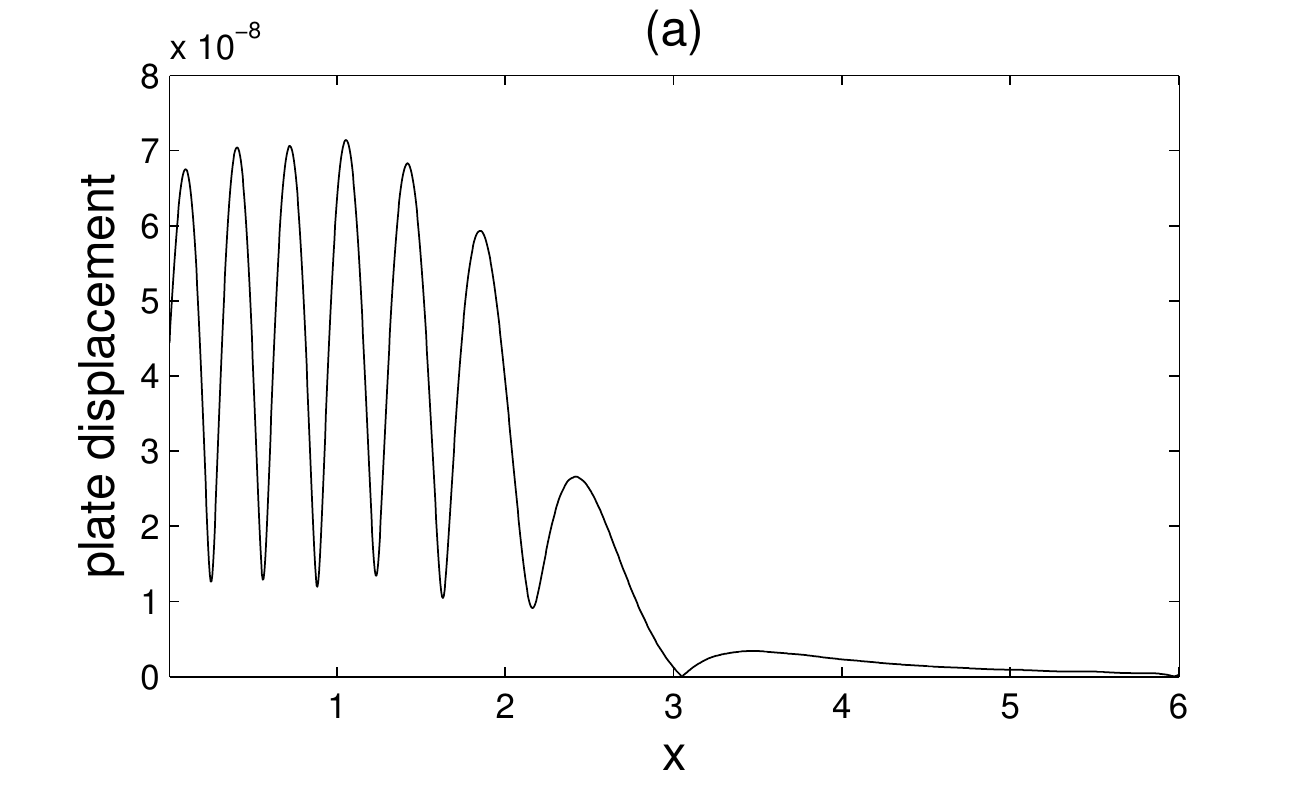}
\includegraphics[width=8cm]
{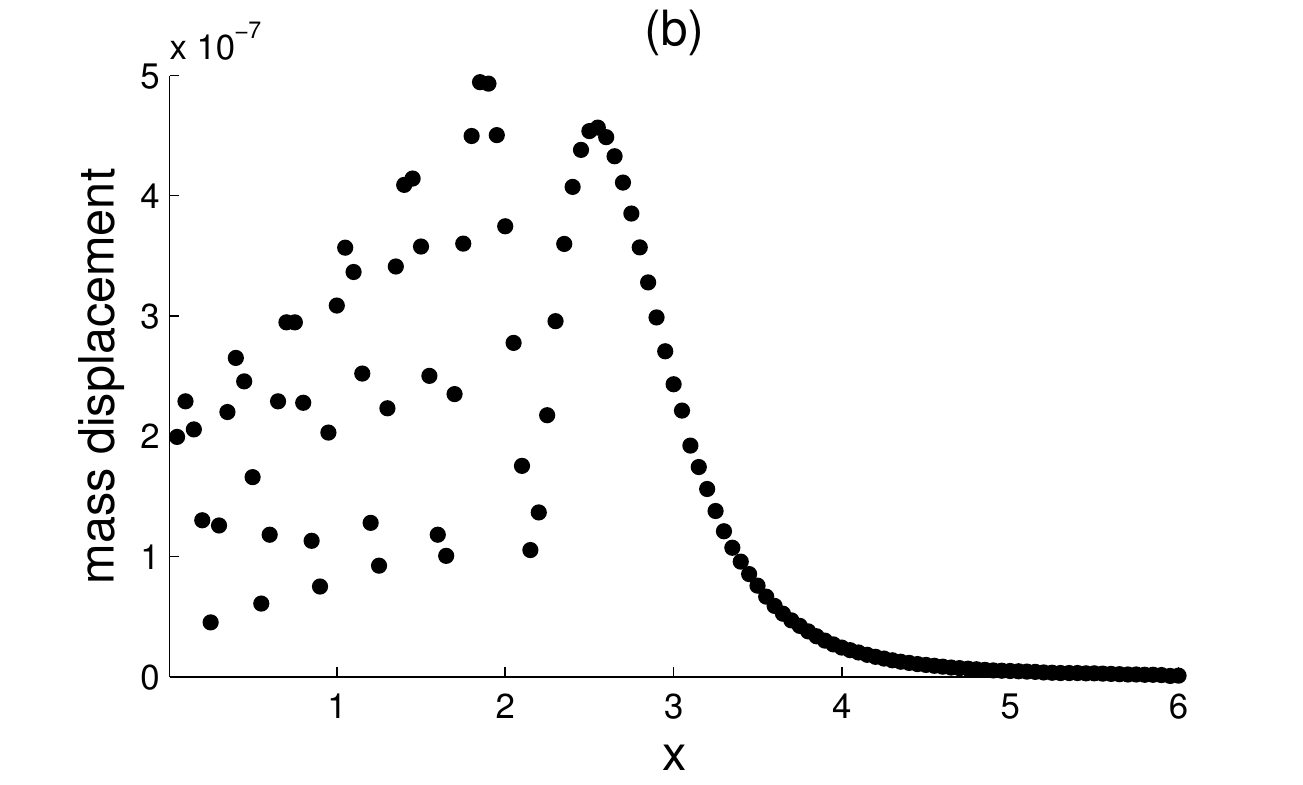}
\caption{Decreasing resonator mass: Time-harmonic (250~Hz)
  displacements of 1cm aluminium plate with water-loading and
  mass-spring resonators with the resonant frequency of mass-spring
  located at $x=3$ at 250~Hz. The absolute values of the plate and
  mass displacements  are shown in the top
and bottom panels respectively.}
\label{fig:RampDownU}
\end{figure}

We first consider what happens to a wave as it moves through an array
of gradually decreasing masses, i.e. ramping down the mass. From
Fig.~\ref{fig:MassDisp.1} for 
fixed frequency we start from the top of the graph and follow
the dispersion curve downwards. Initially the surface wave propagates there with a horizontal
wavenumber in the plate close to that of the surface wave in a pristine
plate,
and in the fluid with corresponding exponential decay in the direction
perpendicular to the plate; 
this is effectively the dashed line of Fig.~\ref{fig:AlphaDisp.1}. As the wave
progresses through the array, it encounters ever smaller values of the resonator mass, and
the wavenumber adapts by also decreasing
c.f. Fig.~\ref{fig:MassDisp.1} and the exponential rate of decay of
the wave in the fluid 
perpendicular to the plate also decreases. The surface wave hybridises
and moves towards the bulk acoustic wave branch 
until, for say 250~Hz at $m \sim 1$, the
curve arrives at the band
gap and meets the acoustic branch. 
At this point the system can no
longer support a wave propagating in the direction parallel to the
plate as the resonator mass decreases further. However the energy
within the surface wave has to go somewhere, and the energy
contained in the wave must be radiated out into the fluid, the point
at which this occurs can be explicitly predicted from Fig. \ref{fig:BandGap}. Thus
we have constructed a situation where mode conversion from a surface
wave to a bulk wave occurs and where, as far as the surface
displacement is concerned, there is both vanishing reflection and
transmission. The cosine of the angle at which waves leave the surface is simply
given by the ratio of sound speeds, i.e. $\cos\theta=(c_f/c_p)$, and,
as shown in, Fig. \ref{fig:anglefreq}, and is verified from the
numerical results. 


If we now consider the contrasting case, that of ramping up the mass
with a wave moving through an array
of gradually increasing masses, that is we start from the lower branch
of the curve, which cuts on at zero mass and $\alpha =\alpha_{1}$. As
this wave progresses it encounters ever larger values of resonator
mass and the wavenumber adapts and increases, until when
$\alpha=\pi/d$ the curve arrives at the lower edge of the band
gap. However, at this point, because of the continuity of the curve
there (due to symmetry and periodicity) the wave is totally
reflected. This is the origin of so-called rainbow trapping as the
position at which this reflection occurs is frequency dependent and so
one can  design selective spatial frequency
 separation of waves.

\begin{figure}  
\includegraphics[width=9.5cm]{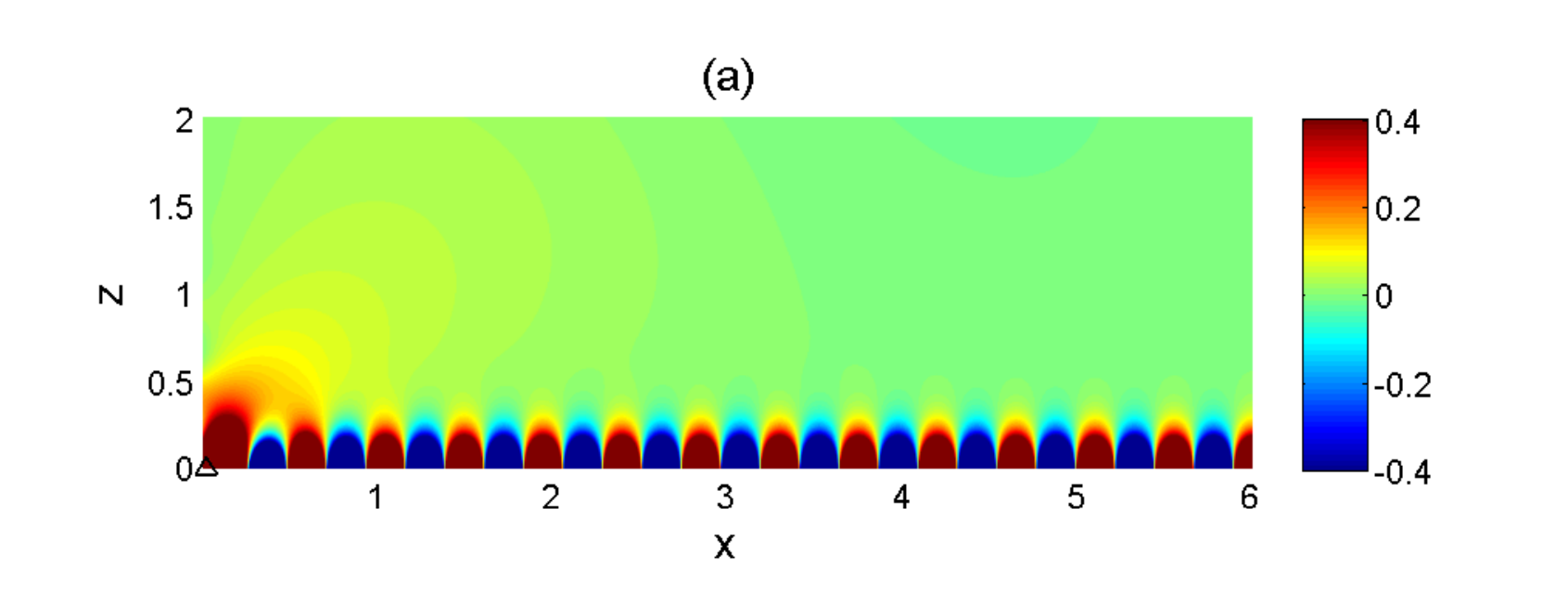}
\includegraphics[width=10cm]{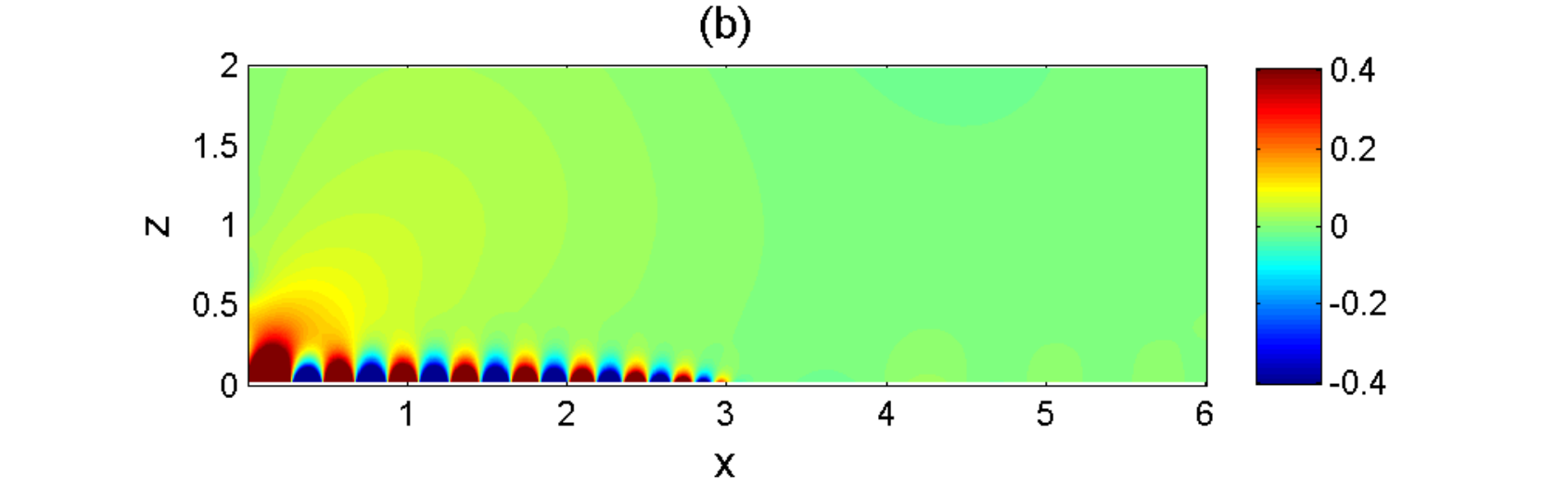}
\caption{Time-harmonic (250Hz) snapshot of pressure in water above 1cm
  aluminium plate: Top, a pristine plate without resonators and
  bottom, 
with mass-spring resonators. Resonator mass
  increases with $x$. Resonant frequency of mass-spring located at $x=3$ is 250Hz. }
\label{fig:RampUpPress}
\end{figure}

\begin{figure}  
\includegraphics[width=8cm]{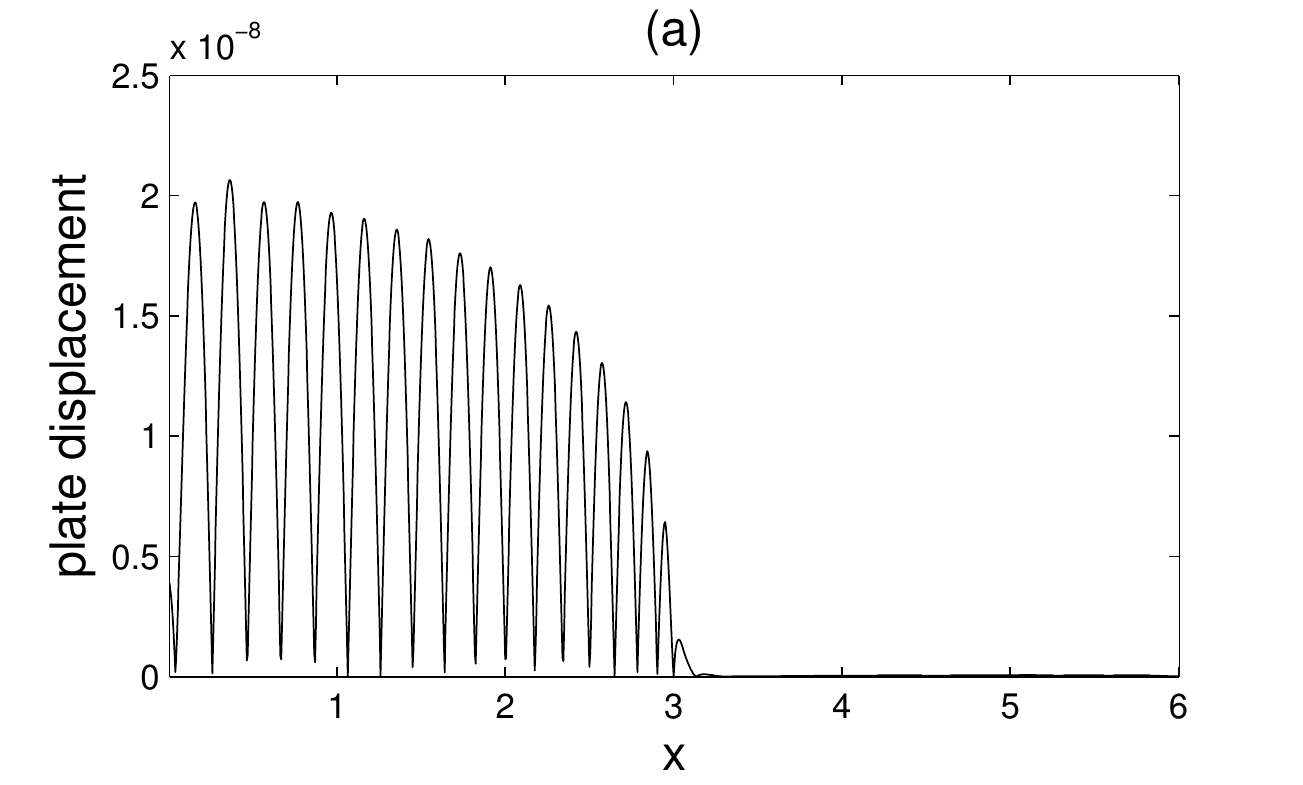}
\includegraphics[width=8cm]{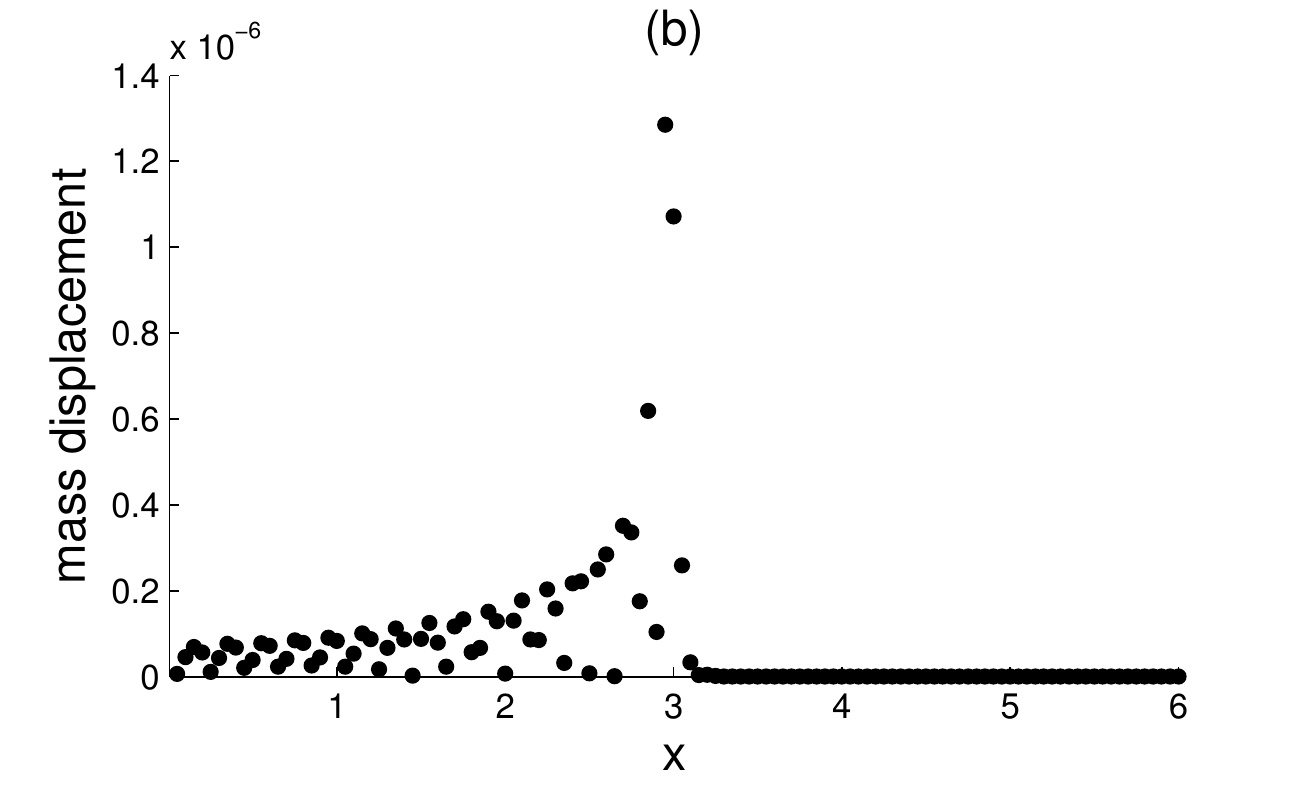}
\caption{Increasing resonator mass: Time-harmonic (250~Hz)
  displacements of 1cm aluminium plate with water-loading and
  mass-spring resonators with the resonant frequency of mass-spring
  located at $x=3$ at 250~Hz. The absolute values of the plate and
  mass displacements,  are shown in the top
and bottom panels respectively.}
\label{fig:RampUpU}
\end{figure}
%


\section{\label{sec:results} Numerical results}%

We now turn to numerical simulations to illustrate the mode conversion
and rainbow trapping effects. The numerical simulations use the
finite difference method together with perfectly matched layers (PML)
\cite{berenger94a,skelton07a} which are used to mimic and infinite
domain; this is done by surrounding a finite area of fluid, and
the corresponding finite length of plate, with a finite thickness of
PML. The forcing is taken to be at the origin and of foem
$\delta(x)exp(-i \omega t)$. 
 This is implemented in finite differences in the frequency domain
 with at least 10 grid points between each resonator  to ensure
 sufficient resolution per wavelength.

We first consider the effect of ramping down the masses and the
potential for mode conversion.  We use 120 mass-spring resonators with spacing $d=0.05$~m and graduate
the masses by linearly ramping them down from 1.3~kg to 0.705~kg in
graduations of 0.005 ~kg. The spring constant $\lambda$ is fixed as
$2.4674\times10^6$ $\mathrm{kg}\cdot \mathrm{s}^{-2}$ which is chosen to give a resonant freq of 250~Hz for $m=1$~kg.
 The mode conversion is strikingly seen in
Fig. \ref{fig:RampDownPress},  which shows the surface wave propagating along the
array with a gradually increasing wavelength, see Fig. \ref{fig:RampDownU},  until it reaches the
critical point, here $x\sim 3$~m, where the surface wave mode has
hybridised into a bulk wave and the energy radiates into the bulk as
highly directional long waves; clearly the resonator spacing (of 0.05~m) is
sub-wavelength and for comparison the wavelength for the unconstrained
fluid-loaded plate is  $0.4508$~m and the wavelength in the fluid $6$~m.
 This striking mode conversion is accompanied by decreasing plate and
resonator displacements as the wave mode coverts.

The rainbow trapping phenomena is illustrated in
Fig. \ref{fig:RampUpPress}, and the masses now increase in
amplitude. This figure also shows the wavefield in the absence of
resonators as a reference for comparison with
Figs. \ref{fig:RampDownPress}, as well as the varying masses  case of
Fig. \ref{fig:RampUpPress}.
We again use 120 mass-spring resonators with spacing
$d=0.05$~m and reverse the graduation
with the masses linearly ramping up from 0.705~kg to 1.3~kg in
increments of 0.005 ~kg and with the spring constant $\lambda$ is fixed as
$2.4674\times 10^6$ $\mathrm{kg}\cdot \mathrm{s}^{-2}$.
The surface wave has gradually decreasing amplitude as it passes
through the array and gets
cut-off at $x\approx 3$;  for different frequencies this position
varies and this position can be read off directly from Fig. \ref{fig:BandGap}.  
Here we witness almost total reflection, and 
notably the plate surface displacement tends to zero, but the
amplitude of the mass resonators increases dramatically, Fig. \ref{fig:RampUpU}, thus the
motion is spatially localised on the resonators which is ideal for applications
such as energy harvesting.

\section{\label{sec:3}  Discussion}
 
We have illustrated that metamaterial ideas, that is using arrays of
local sub-wavelength structuration to influence global wave behaviour,
are relevant to fluid-loaded structures. We have concentrated upon a
feature, surface graded arrays of resonators, pertinent to energy
harvesting, structural vibration damping, mode conversion and the control of waves on surfaces. We
envisage, as in the theory of surface elastic wave mode conversion and
rainbow trapping \cite{daniel17a} which prompted experimental
verification \cite{colombi17a}, that the theory we present here will
drive experimental work. In more general terms, 
 it is clear that other features such as effective negative parameters,
analogies of transformation optics and much more are transferable from
metamaterials to
the field of structural acoustics. It is also important to note that
we have treated the simplest possible type of resonator, masses and
springs, but the analysis easily generalises to more realistic
rod-resonator arrays as utilised in, the non-fluid loaded plate,
analyses of \cite{williams,daniel17a}. 

We also remark that, in recent years, there has been significant interest in so-called \emph{bifunctional metamaterials} which aim to harness the power of two different physical phenomena to achieve metamaterial effects.
Thus far however, these approaches have focused on the control of fields that, fundamentally, are governed by the same physics and have the same underlying mathematical structure (see, for example,~\cite{kippenberg2007cavity,rolland2012acousto,moccia2014independent,lan2016bifunctional}); thermal and electrical conduction are both governed by the Helmholtz operator, for example.
In contrast, the present paper represents a step change in the study of \emph{bifunctional metamaterials}.
Here we demonstrate a truly bifunctional metamaterial, synthesising two fundamentally different physical systems, acoustic fluid (governed by the Helmholtz operator) and flexural plate waves (governed by the biharmonic operator), to create the striking mode conversion illustrated in \S\ref{sec:Variable}.

We emphasise that, whilst motivated by the the seismic metawedge~\cite{colombi16a,daniel17a}, the present work goes far beyond the mode conversion demonstrated by the metawedge.
The elastic metawedge examined in~\cite{colombi16a,daniel17a} is capable of converting elastic surface waves into elastic bulk waves, that is, conversion from one elastic mode to another.
In contrast, the fluid-loaded metasurface is capable of converting flexural plate waves to acoustic waves, that is, converting mechanical waves into acoustic waves.
In this sense, the fluid-loaded metasurface studied in the present paper transcends the boundaries between disparate physical phenomena creating a truly bifunctional metasurface. 


\begin{acknowledgments}
This research was supported by the U.K. EPSRC under grant
EP/L024926/1 and by a Marie-Curie Fellowship to AC. 
\end{acknowledgments}



\end{space}

\bibliographystyle{jasa-ml}
\section*{References}




\end{document}